\documentclass[prb,amsmath,amssymb]{revtex4}

\begin{document}


\title{The discretised harmonic oscillator: Mathieu functions and
a new class of generalised Hermite polynomials}


\author{M. Aunola}
\email[email: ]{Matias.Aunola@phys.jyu.fi}
\affiliation{Department of Physics, University of Jyv\"askyl\"a,
P.O. Box 35 (YFL), FIN-40014 University of Jyv\"askyl\"a, FINLAND}



\date{\today}

\begin{abstract}
We present a general, asymptotical solution for the 
discretised harmonic oscillator. The corresponding
Schr\"odinger equation is canonically conjugate to the 
Mathieu differential equation, the Schr\"odinger equation
of the quantum pendulum. Thus, in addition to
giving an explicit solution for the Hamiltonian of
an isolated Josephon junction or a superconducting 
single-electron transistor (SSET), we obtain an 
asymptotical representation of Mathieu functions.
We solve the discretised harmonic oscillator by transforming
the infinite-dimensional matrix-eigenvalue problem 
into an infinite set of algebraic equations which
are later shown to be satisfied by the obtained solution.
The proposed ansatz defines a new class of
generalised Hermite polynomials which are explicit functions
of the coupling parameter and tend to ordinary Hermite polynomials
in the limit of vanishing coupling constant. 
The polynomials become orthogonal as parts of the eigenvectors 
of a Hermitian matrix and, consequently, the
exponential part of the solution can
not be excluded. We have conjectured the general structure
of the solution, both with respect to the quantum number 
and the order of the expansion. An explicit proof is given
for the three leading orders of the asymptotical solution
and we sketch a proof for the asymptotical convergence
of eigenvectors with respect to norm. 
From a more practical point of
view, we can estimate the required effort for improving the 
known solution and the accuracy of the eigenvectors. The applied
method can be generalised in order to accommodate several variables.
\end{abstract}


\maketitle


\section{Introduction\label{sec:intro}}

This paper is closely related to one of the famous eigenvalue problems,
namely that of a one-dimensional harmonic oscillator. 
It is common knowledge that if the eigenvectors
are required to have continuous  second-order derivatives, each eigenvector
is expressible as a product of a Hermite polynomial and an exponential
term. The corresponding eigenvalues are equidistantly spaced and bounded
from below. Another way to state the problem is given by the 
annihilation and creation operators which directly  
diagonalise the Hamiltonian. In comparison, the quartic anharmonic
oscillator was solved by Bender and Wu in Ref.~\onlinecite{ben69}.
A method for finding eigenvalues for 
anharmonic oscillators was created by Mei\ss ner
and Steinborn in Ref.~\onlinecite{mei97}. A general
method for polynomial potentials was introduced
recently by Meurice.~\cite{meu02,bac95}

Instead of continuous functions, we consider functions defined
only on a discrete, equidistantly-spaced and countable 
set on $\mathbb{R}$. 
The obvious advantage of this approach is that it transforms
the problem into an eigenvalue problem of 
an infinite-dimensional, tri-diagonal matrix. The corresponding 
Schr\"odinger equation is canonically conjugate to 
the Mathieu differential equation.\cite{abr72} Numerical solutions
for noninteger orders are naturally obtained by diagonalising the
very same matrix, see Ref.~\onlinecite{shir93} and the references
therein for applications. 

In physics, the discretised harmonic oscillator is manifestly 
realised by the Hamiltonian of an isolated Josephson 
junction\cite{cel00,tink96} and the Hamiltonian
of the, slightly misleadingly named,
superconducting single-electron transistor 
(SSET).\cite{eil94,tink96} Presently, excited states are seldom
considered because of the radical approximations under which 
the Hamiltonian is solved. Even if the excited states
are numerically obtained,
it is not immediately evident, what happens when the coupling is 
changed. In this article we give an explicit, asymptotical
solution for the discretised harmonic oscillator which corresponds
to strong Josephson coupling in case of the SSET. The same 
Hamiltonian also describes the so-called quantum pendulum, or
a particle in a periodic potential.\cite{dor02,sto78}

The corresponding asymptotical eigenvalues have 
been available for almost
fifty years due to the work of Meixner and Sch\"afke on 
Mathieu functions in
Ref.~\onlinecite{mei54}. 
First, by calculating the 
determinant of the matrix representation accurately 
enough, we continue this expansion
by several orders in the coupling parameter.
Then we propose an ansatz that transforms the matrix
equation into an infinite set of algebraic equations
and proceed by recursively solving these equations.
The general properties of the coefficients in the ansatz
can be obtained by studying any occurring regularities and 
reinserting these into the solution. Thus, in addition to
the eigenvalues, we have successfully conjuctured the 
general form of the asymptotical eigenvectors. 
In each order of the expansion, the expressions are quoted in
terms of an arbitrary quantum number, $n$, whenever possible.
The leading terms have been determined and rephrased in terms
of an arbitrary order, $m$, too. We find that the eigenvectors
are asymptotical solutions of certain differential equations,
which enables us to obtain further orders in their expansions.

The only real-valued parameter in the 
solution is the coupling constant, because all coefficients, 
both in the eigenvalues and in the ansatz are rational
numbers. As a  practical application,
the rate of convergence of the solution towards 
numerically obtained, ''exact'', solution, can be
reliably estimated. In the asymptotical limit, the
dependence in terms of $n$ and $m$ assumes the form
of a simple monomial, at least down to the limits of 
numerical precision. 

The solutions of order $m\le 5$ are very simple to 
program and directly apply as numerical solutions of the
discretised harmonic oscillator. For sufficiently
small values of the coupling constant the eigenvectors
are practically exact and thus they facilitate 
studies which require the structure of the excited states.
We have proven, with the help of recursion relations of
Hermite polynomials, that the first three leading orders
of the obtained solution are correct. The calculation
up to the seventh order should be performed in the future.
We also outline an explicit proof concerning
the normwise convergence of the eigenvectors. The asymptotical
nature of the solutions must be stressed. A very thorough
introduction on the subject has been given been given by
Boyd in Ref.~\onlinecite{boy99}. 

It is justified to ask, is the proposed solution completely
new. The answer is, naturally, yes and no. Both discretised
and discrete harmonic oscillators have been widely studied
before. Both cases are related to orthogonal polynomials,
so the work of Kravchuk\cite{kra29} and Hahn\cite{hah49} 
must be mentioned. The discrete harmonic oscillator, where 
the position 
coordinate is restricted to a finite number of values,
is explicitly solved by Kravchuk polynomials as shown by
Lorente in Ref.~\onlinecite{lor01}. Several discretisations
of the harmonic oscillator have been previously solved, each
giving rise to a specific class of generalised Hermite 
polynomials. Discretisation by an exponential 
lattice $\{-q^n,q^n\vert n\in\mathbb{Z}\}$, where $0<q<1$, 
defines the so-called $q$-deformed harmonic oscillator and
generalised $q$-Hermite polynomials which are rigorously 
discussed by Berg and Ruffing in Ref.~\onlinecite{ber01}. 
For other applications of the $q$-deformed harmonic oscillators,
see e.g. Refs.~\onlinecite{par94} and \onlinecite{bon94}, where
other discretisations are reviewed, too.  Borzov, in 
Ref.~\onlinecite{bor01}, considers generalised derivation 
operators as generators of Hermite polynomials and 
states that the generalised
Hermite polynomials either satisfy a second order differential
operator or there is no differential equation of finite order
for these polynomials. Many other types of generalisations
are also known, see e.g. the multi-dimensional  Hermite 
polynomials of  R\"osler\cite{ros98}, Hermite polynomials
orthogonal with respect to the measure $\vert \xi\vert^\gamma
\exp(-\xi^2)d\xi$, where $\gamma>-1$\,\cite{ros93,det96}, and
parabosonic Hermite polynomials.\cite{jin02} 
In the future, it must be established whether the
presented class of Hermite polynomials is related to 
the $q$-Hermite polynomials, if it results from some 
other discretisation or is it an explicit 
example of the second group of Borzov's categorisation. 
Complementary
results concerning the introduction of distant boundaries
for the continuous problem are also known.\cite{bar90,eli99}
Finally, it should be
emphasised that instead of deforming the harmonic oscillator,
we solve its common-sense discretisation, used especially in
numerical calculations. The asymptotical effects of the 
discretisation are explicitly calculated.

We also briefly consider the 
abruptly changing nature of the solutions when the
coupling constant vanishes. This behaviour is evident
for both versions of the harmonic oscillator and 
the Mathieu  differential equation. The asymptotical
nature of the solutions and the eigenvalues is 
caused by this divergence.  For the Mathieu equation 
this has been well documented, see e.g. 
Refs.~\onlinecite{mei54,mei80,abr72}.
A more physically motivated approach is given by  
Bender, Pelster and Weissbach in Ref.~\onlinecite{ben02},
where e.g. the instanton equation and the Blasius equation
are examined. The present methods are closely related to
these, although we can not carry the calculation 
as far in the perturbative expansion. 
This is explained by the necessity
of obtaining the expansion for the eigenvalues which 
makes the present problem technically more demanding. 

The present method can be generalised in a fairly
obvious manner.
Other differential equations with analytical solutions 
can be discretised in the same manner if the correct
expansions are found for all parts of the solution.
An easier generalisation is related to multi-dimensional 
difference equations with harmonic (quadratic) potential 
terms. The existing solution\cite{aun02} for Hamiltonians of  
one-dimensional arrays of Josephson junctions become more
transparent with the help of present formalism.

The paper is organised as follows. In 
Sec.~\ref{sec:discretised} we define the discretised 
harmonic oscillator and connect it to the 
Mathieu differential equation as well as the continuous
case. The solution ansatz and the resulting set of 
equations are reviewed.  In Sec.~\ref{sec:solu} we quote
our conjectures for the general form of the coefficients in 
the ansatz. We also present the explicit values of 
the leading coefficients.  In Sec.~\ref{sec:ansatz}
we study solving the set of equations which yields 
the asymptotical eigenvectors. Efficient truncations
of the set of equations are explained. The effort 
for improving the obtained results with the present 
method is estimated. In Sec.~\ref{sec:proof}  
we prove that the solution satisfies the difference equations,
at least for the three leading orders. The rate of convergence
and the induced asymptotical orthonormality are also reviewed.
Finally, in Sec.~\ref{sec:conclu} the conclusions are drawn and
an outlook of future possibilities is given.   

A final note for those that are only interested in applying these 
results in numerical and/or theoretical analysis. Please review the
beginning of Sec.~\ref{sec:discretised} in order to find the 
correct parameters for the discretised harmonic or Mathieu 
equation. Then proceed to Sec.~\ref{sec:solu} and use the 
given expressions as approximate solutions in 
Eq.~(\ref{eq:genansatz}).   

\section{The discretised harmonic oscillator
\label{sec:discretised}}

The eigenvalue problem corresponding to the harmonic oscillator
is the differential equation for $\psi(x)$,
\begin{equation}
-\frac12\frac{d^2\psi}{dx^2}+\frac{\omega^2 x^2}{2}\psi=\lambda\psi.
\label{eq:diff0}
\end{equation}
The eigenvectors corresponding to the well-known eigenvalues,
\begin{equation}
\lambda_n=\omega(n+1/2),
\end{equation}
where $n=0,1,2,\ldots$, are given by
\begin{equation}
\psi_n(x)=A_nH_n(\xi)e^{-\xi^2/2}.\label{eq:contsolu}
\end{equation}
Here $\xi=\sqrt{\omega}\,x$, 
$A_n$ is a normalisation factor, and $H_n$ is the Hermite 
polynomial of order $n$.
The Hermite polynomials are solutions of the
Hermite differential equation
\begin{equation}
y''-2xy'+2ny=0,
\end{equation}
where $n=0,1,2,\ldots$. For our convenience, 
we write the polynomials, given by Rodrigues' formula, as 
\begin{equation}
H_n(\xi)=(-1)^n\exp(\xi^2)\frac{d^n}{d\xi^n}\exp(-\xi^2)
=\sum_{k=0}^{k'}h^{(n)}_k\xi^{n+2(k-k')},\label{eq:hermitepol}
\end{equation}
where $k':=\lfloor n/2\rfloor$, i.e. 
$k'=n/2$ if $n$ is even and $k'=(n-1)/2$\ if $n$ is odd. 
The quantity $k'$ proves to be extremely useful in further
analysis. The Hermite polynomials satisfy the recursion relation
\begin{equation}
H_{n+1}(\xi)=2\xi H_n(\xi)-2n H_{n-1}(\xi).\label{eq:hermrecur}
\end{equation}
Many of the generalisations of the Hermite polynomials boil 
down to a generalisation of this recursion 
relation.\cite{lor01,ber01,bon94,jin02}  

The discretised version of Eq.~(\ref{eq:diff0}) is 
obtained by restricting the values 
of $x$ onto an evenly spaced, countable subset of $\mathbb{R}$. This
corresponds e.g. to the discretisation of charge in case of a
Josephson junction or a SSET. 
Only the constant nearest-neighbour coupling is retained which yields 
a tri-diagonal matrix $H(x_0)$ with non-zero matrix elements  
\begin{equation}
H_{jj}(x_0)=\mbox{$\frac12$}\omega^2(j-x_0)^2,\quad H_{j+1,j}(x_0)=H_{j,j+1}
(x_0)=-\mbox{$\frac12$}.\label{eq:discrete}
\end{equation}
Here the parameter $x_0\in[-\mbox{$\frac12$},\mbox{$\frac12$}]$ 
is the displacement of the origin 
with respect to the matrix element $j=0$. All eigenvalues of $H(x_0)$
have been translated by $-1$ in order to simplify the diagonal
matrix elements. The standard way to write the Hamiltonian 
of an inhomogeneous SSET is obtained from Eqs.~(7.36) and (7.39)
of Ref.~\onlinecite{tink96} and rephrasing it in terms of the
number operator for Cooper pairs yields the matrix 
\begin{equation}
H^{(\mathrm{SSET})}_{jj}(N_0)=E_{\mathrm{C}}(j-N_0)^2,
\quad H^{(\mathrm{SSET})}_{j+1,j}(N_0)=H^{(\mathrm{SSET})}_{j,j+1}
(x_0)=-\mbox{$\frac12$}E_{\mathrm{J}}(\theta),
\end{equation}
where $N_0$ is the number of Cooper pairs which minimises the charging
energy, $E_{\mathrm{C}}=(2e)^2/2C_{\Sigma}$ is the unit of charging 
energy, and  $E_{\mathrm{J}}(\theta)$ is the effective Josephson energy
which depends on the total phase $\theta$ 
across the SSET. Consequently, we
solve the Hamiltonian of SSET if we find the eigenenergies and
eigenvector for the discretised harmonic oscillator with
$\omega=(2E_{\mathrm{C}}/E_{\mathrm{J}}(\theta))^{1/2}$.   

In the following, we are searching for 
eigenvectors with finite 
Euclidean norm, i.e. 
\begin{equation}
\Vert\psi\Vert^2=\sum_{j=-\infty}^\infty\vert 
\psi_j\vert^2<\infty.
\end{equation}
The existence and uniqueness of such solutions 
follows from the generalisation of the Gershgorin eigenvalue theory
by Shivakumar, Rudraiah and Williams in Ref.~\onlinecite{shiv87}.  
First the number of eigenvalues of $H(x_0)$ 
on a given interval can be shown to coincide with 
number of eigenvalues for a finite-dimensional truncation
of the matrix, $H^{(N)}(x_0)$, if the dimension $N$ is sufficiently
large. A sufficient condition for this is that the
difference between ordered diagonal elements exceeds 
$2\times\vert-\frac12\vert=1$.
When $x\approx 0.5$, this property is obtained more easily for 
even values of $N$.
Furthermore, they prove that, for finite values of $n$, 
the eigenvector $\psi^{(N)}_n$ of $H^{(N)}(x_0)$ tends to the 
corresponding eigenvector of $H(x_0)$ 
when $N\rightarrow \infty$.

We now establish the connection between the $H(x_0)$ and the Mathieu
differential equation\cite{abr72}
\begin{equation}
\frac{d^2y}{dv^2}+(a-2q\cos(2v))y=0,\label{eq:canoneq}
\end{equation}
where $a$ is the eigenvalue, also known as the 
characteristic value when the solution $y$ has period of
$\pi$ or $2\pi$.  We follow the derivation of Shirts in
Ref.~\onlinecite{shir93} and use Floquet's theorem to obtain
\begin{equation}
y=\exp(i\nu v)P(v)=\exp(i\nu v)\sum_{k}c_{2k}\exp(2i k v)\label{eq:mathfunc}
\end{equation}
where the Fourier expansion of $P(v)$ has been inserted.
This corresponds to the matrix equation for the coefficients $c_{2k}$
compactly written as
\begin{equation}
c_{2k-2}-V_{2k}c_{2k}+c_{2k+2}=0,
\end{equation}
where $V_{2k}=[a-(\nu+2k)^2]/q$. This is identical to the discretised
harmonic oscillator Hamiltonian $H(x_0)$ with an eigenvalue $\lambda$
after identifications
\begin{equation}
\nu=2x_0,\quad k=j,\quad q=4/\omega^2,\quad a=8\lambda/\omega^2.
\label{eq:mathparam}
\end{equation}
Thus all results obtained for the discretised harmonic oscillator
also hold for Mathieu functions~(\ref{eq:mathfunc}) with 
parameters given in Eq.~(\ref{eq:mathparam}). For $x_0=0$ 
and $x_0=\pm \mbox{$\frac12$}$ the solutions of $H(x_0)$ can be 
chosen to be even or odd with respect to $j$. This corresponds to
writing $P(v)$ in terms of sines and cosines. Special attention
must be given to the even solutions of  $H(x_0=0)$, where the 
resulting equations in the matrix representation read 
\begin{eqnarray}
-\psi_{1}/\sqrt{2}&=&\lambda_{2n} \psi_0,\\
-\psi_0/\sqrt 2+\omega^2 \psi_1/2-\psi_{2}/2&=&\lambda_{2n}\psi_1,\\
-\psi_{j-1}/2+\omega^2j^2\psi_j/2-\psi_{j+1}/2&=&\lambda_{2n}\psi_j.\quad
j\ge 2.
\end{eqnarray} 
The eigenvalues for $x_0=0$ correspond to characteristic
values $\{a_{2n}(q),b_{2n}(q)\}$, while the case 
$x_0=\pm \mbox{$\frac12$}$ is linked to $\{a_{2n+1}(q),b_{2n+1}(q)\}$
as defined in Ref.~\onlinecite{abr72}.  

The asymptotical expansion of the eigenvalues 
corresponding to the limit
$q\rightarrow \infty$ or $\omega\rightarrow 0$ was obtained 
by Meixner and Sch\"afke in Ref.~\onlinecite{mei54}.
The derivation of the eigenvalues is based
on the three-term recurrence relations  
for the Mathieu functions and the requirement that the
norm of the error in the eigenvalue equation 
vanishes faster than a specific  power of $\omega$. 
Meixner and Sch\"afke
quote the asymptotical characteristic values of the Mathieu
equation up to and including the order $\omega^7$ in
Theorem 7 in Sec.~2.3.
Some error estimates for asymptotical expansions of Mathieu functions 
by M. Kurz are given in Ref.~\onlinecite{mei80}. Because the
Mathieu equation is also the Schr\"odinger equation of the
quantum pendulum or a particle in a periodic potential, it
has been studied independently in physics, too.\cite{din62,sto78,dor02}
Especially, the same general expansion for eigenvalues and several
further terms for the ground state energy were obtained by Stone and Reeve
in Ref.~\onlinecite{sto78}.    

In this limit, we can write the eigenvalues of $H(x_0)$ as
\begin{equation}
\lambda_{n}\sim\sum_{m=0}^\infty \lambda_{n}^{(m)}\omega^m, 
\end{equation}
where $\omega\rightarrow 0$, and
\begin{equation}
\lambda_{n}^{(m)}=\sum_{k=0}^{m'}\lambda^{(m)}_{n,k}\hat n^{m+2(k-m')}
\label{eq:hermeig}
\end{equation}
with $\hat n:=2n+1$ and  $m'=\lfloor m/2\rfloor$.
This structure is identical to that of the  Hermite 
polynomials~(\ref{eq:hermitepol}), if one identifies $\hat n$ 
with $\xi$.  
By Ref.~\onlinecite{mei54}, the eigenvalues~(\ref{eq:eigenexp}) 
do depend on $x_0$, but this dependence decreases exponentially
as $\omega\rightarrow 0$.
The maximal difference is given by\cite{mei54}
\begin{equation}
\lambda_n(x_0=\pm \mbox{$\frac12$})-\lambda_n(x_0=0)\sim (-1)^n B_0(1-B_1\omega )
\omega^{-n-3/2}\exp(-8/\omega),
\end{equation}
where $B_0$ and $B_1$ depend on $n$ but not on $\omega$.

This allows us to write the eigenvalues of $H(x_0)$ as
\begin{eqnarray}
\lambda_n&\sim&-1+\frac{\omega\hat n}2-\frac{\omega^2d_2}{2^6}-
\frac{\omega^3d_3}{2^{11}}-
\frac{\omega^4d_4}{2^{17}}-\frac{\omega^5d_5}{2^{23}}
-\frac{\omega^6d_6}{2^{27}}
-\frac{\omega^7d_7}{2^{33}}-\frac{\omega^8d_8}{2^{40}}-
\frac{\omega^9d_9}{2^{47}}-\cr
&&\ \frac{\omega^{10}d_{10}}{2^{51}}-\frac{\omega^{11}d_{11}}{2^{57}}
-\frac{\omega^{12}d_{12}}{2^{61}}-\frac{\omega^{13}d_{13}}{2^{69}}
-\frac{\omega^{14}d_{14}}{2^{72}}-\frac{\omega^{15}d_{15}}{2^{79}}
-\frac{\omega^{16}d_{16}}{2^{87}}+\mathcal{O}(\omega^{17}).\label{eq:eigenexp}
\end{eqnarray}
where the coefficients $d_k$ read
\begin{eqnarray}
d_2&=&\hat n^2+1,\cr
d_3&=&\hat n^3+3\hat n,\cr
d_4&=&5\hat n^4+34\hat n^2+9,\cr
d_5&=&33\hat n^5+410\hat n^3+405\hat n,\cr
d_6&=&63\hat n^6+1260\hat n^4+2943\hat n^2+486,\cr
d_7&=&527\hat n^7+15617\hat n^5+69001\hat n^3+41607\hat n,\cr
d_8&=&9387\hat n^8+388780\hat n^6+2845898\hat n^4+ 
  4021884\hat n^2+506979,\cr
d_9&=&175045\hat n^9+9702612\hat n^7+107798166\hat n^5+
   288161796\hat n^3+130610637\hat n,\cr
d_{10}&=&422565\hat n^{10}+30315780\hat n^8+480439190\hat n^6+
   2135766820\hat n^4+2249346285\hat n^2+238353840,\cr
d_{11}&=&4194753\hat n^{11}+379291385\hat n^9+8186829426\hat n^7+
   55529955498\hat n^5+110241863469\hat n^3+\cr
   &&\ 41540033277\hat n.\cr
d_{12}&=& 10645960\hat n^{12} + 1187264199\hat n^{10} + 33678377895\hat n^8 + 
  327725946398\hat n^6+\cr 
  &&\ 1081358909790\hat n^4 +940077055035\hat n^2 + 88258370067\cr
d_{13}&=& 440374207\hat n^{13} + 59495737574\hat n^{11} 
   + 2155821044201\hat n^9 + 
    28738150160500\hat n^7 +\cr
   &&\  144821249264769\hat n^5 +236410740537606\hat n^3
     + 78243613727607\hat n\cr
d_{14}&=& 578183175\hat n^{14} + 93209584104\hat n^{12}
  + 4215683624295\hat n^{10} +
    74269604367684\hat n^8 +\cr 
  &&\ 537905750769429\hat n^6 + 
	  1456767306013752\hat n^4 +1105711550410653\hat n^2 + 94839535889532\cr
d_{15}&=& 12308013927\hat n^{15} + 2337227706555\hat n^{13} 
  + 129437253243675\hat n^{11} + 
  2928506455684095\hat n^9 +\cr
  &&\  29119560960614085\hat n^7 + 120372998803922241\hat n^5 + 
  170921920649402745\hat n^3 +\cr
  &&\  51316344023990085\hat n\cr
d_{16}&=& 530039126159\hat n^{16} + 117243302735480\hat n^{14} 
 + 7823093961425652\hat n^{12} + 
  222043810819026856\hat n^{10}+\cr
  &&\ 2924952921130025194\hat n^8 + 17380315268028265224\hat n^6 + 
  40851669411526600980\hat n^4 +\cr
  &&\  27983551470330365784\hat n^2 + 2235152520630714879\nonumber
\end{eqnarray}
We obtain the terms for orders $8\le m\le 11$   
by exploiting Eq.~(\ref{eq:hermeig}) when
explicitly evaluating the determinant of Eq.~(\ref{eq:discrete}).
As a first step, setting $x_0=0$ halves the dimension
of the tridiagonal matrix. Next, by translating  one of the eigenvalues 
close to zero by substracting the known expansion of this eigenvalue, 
the  determinant becomes an essentially linear  function of the 
chosen, translated  eigenvalue. The next unknown term is inserted as a parameter
and the determinant  is calculated for several values 
of $\omega$, preferably in the form $\{2^{-k}\omega_0\}_{k=0}^{3\ 
\mathrm{to}\ 5}$. This choice lets us separate the leading
correction and the subsequent corrections. In order
to obtain the terms $d_{2k}$ and $d_{2k+1}$, we must 
correctly determine all eigenvalues $\lambda_n$ when $n\le k$.
Sufficient accuracy is guaranteed by using 
the high-precision numerics of \textsc{Mathematica} software.
The method for obtaining the orders $m> 11$ requires explicit
knowledge on the properties of the eigenvectors and 
the discussion is postponed until
the end of Sec.~\ref{sec:solu}.

The asymptotical nature of the expansion means that for
each value of $\omega$ and $n$, there exists and optimal order
$m$ which minimises the error in the eigenvalue, i.e. the function
\begin{equation}
\Delta \lambda(\omega,n,m):=
\left\vert \lambda_n-\sum_{m'=0}^m \lambda_{n}^{(m')}\omega^{m'}
\right\vert,
\end{equation}
with respect to $m$. The exact eigenvalue $\lambda_n$ exists and is finite
for all non-zero values of $\omega$ according to the Sturmian theory of 
second-order linear differential equations, see e.g. 
Ref.~\onlinecite{abr72}. 
In order words, for sufficiently small values of 
$\omega$, $n$ and $m$, the error is dominated by the first omitted
term, i.e. $\Delta \lambda(\omega,n,m)\sim 
\vert\lambda_{n}^{(m+1)}\vert\omega^{m+1}$. 
Because the asymptotical eigenvalue is divergent, it surely crosses
the exact eigenvalue when $\omega$ is increased, but this occurs
outside the range of asymptotical convergence.
Similar asymptotical convergence should be observed
for the asymptotical eigenvectors, too. 
Assuming $\psi_n^{(m,x_0)}$ 
corresponds to the asymptotical expansion of the eigenvalues up 
to and including 
order $\omega^m$,  we expect error in the norm  to behave as
\begin{equation}
\Vert \psi_n^{(m,x_0)}-\psi_n^{(x_0)}\Vert\sim C(n,m)\omega^m,
\label{eq:asymp}
\end{equation}
where $\omega\rightarrow 0$ and 
$C(n,m)$ is a simple function of $n$ and $m$.  Although
this has not been proven, Eq.~(\ref{eq:asymp})
appears to be correct and we will ultimately give an approximate
expression for $C(n,m)$, too. Outside the regime of asymptotical
convergence the error~(\ref{eq:asymp}) approaches $\sqrt 2$ as
the asymptotical solution becomes orthogonal to the exact one.

Next we show that the discrete eigenvalue problem 
Eq.~(\ref{eq:discrete}) is a meaningful asymptotical 
limit of the continuous harmonic oscillator equation~(\ref{eq:diff0}).
The problems are identical in the leading infinitesimal order 
when $\omega$ is infinitesimal, but the limit $\omega\rightarrow 0$ 
is subtle. As long as $\omega>0$, both the eigenvalues 
and eigenvectors of the discretised problem tend to those 
of the continuous harmonic oscillator with this $\omega$.
For $\omega=0$ the continuous problem becomes abruptly
the free particle Hamiltonian with solutions 
\begin{equation}
\psi_{\omega=0}(x)=e^{ikx},\quad \lambda_{\omega=0}=k^2/2,
\label{eq:contomega}
\end{equation} 
where $k$ is the standard name for the wave number.
Simultaneously the discretised problem becomes the well-known 
nearest-neighbour chain with eigenvectors and eigenvalues
\begin{equation}
\psi_k=\{e^{ik(j-x_0)}\}_j,\quad \lambda_{\omega=0}=-\cos(k).
\end{equation} 
For sufficiently small values of $k$ we have $\lambda_{\omega=0}
\approx -1+k^2/2$, in agreement with Eq.~(\ref{eq:contomega}).
In contrast, we are interested
in the bound-state solutions of  Eq.~(\ref{eq:diff0})
and those eigenvectors of the discretised problem that can be
uniquely related to these continuous solutions for $\omega>0$.

The harmonic oscillator is discretised by restricting 
the values of $x$ onto a countable and evenly
spaced subset of $\mathbb{R}$. The lowest-order central approximation 
for a second-order derivative is simply
\begin{equation}
\psi''(x)=\frac{\psi(x-h)-2\psi(x)+\psi(x+h)}{h^2}\ \left[+\mathcal{O}(h^4)
\right]. 
\end{equation}
Assuming $\psi(x)$ to be real-analytic
allows us to write the numerator of the right-hand-side as a Taylor series
\begin{equation}
\psi(x+h)-2\psi(x)+\psi(x-h)=\sum_{k=1}^\infty\frac{2h^{2k}}{(2k)!}
\frac{d^{2k}\psi(x)}{dx^{2k}}.\label{eq:derivatives}
\end{equation}
If $h$ is infinitesimal and as the derivatives of $\psi$ are finite
in all orders,
the only remaining term is $h^2\psi''(x)$. Thus, in the lowest 
infinitesimal order the discretised eigenvalue problem gives a
second order differential equation
\begin{equation}
-\frac12\frac{d^2\psi_x}{dx^2}+\frac{\omega^2x^2}{2h^2}\psi_x=
h^{-2}(-1+\lambda)\psi_x\label{eq:first}
\end{equation}
which is identical to Eq.~(\ref{eq:diff0}) apart from the constant
$-h^{-2}$ and the redefinitions $\omega\mapsto \omega/h$ and
$\lambda\mapsto\lambda/h^2$.
The discreteness of the problem can also be varied by rescaling the
value of $\omega$. Thus, instead of decreasing the size $h$ of the 
steps, we set $h=1$ and let $\omega\rightarrow 0$. From 
Eq.~(\ref{eq:first}) we see that asymptotically the
eigenvalues and eigenvectors have the form
$\lambda_n\sim -1+\omega(n+1/2)$ and $\psi_x\sim \psi(x)$, 
as expected.

We have already pointed out that the matrix $H(x_0)$
in Eq.~(\ref{eq:discrete})  can be derived
from the Mathieu equation. The underlying reason
for this is that the problems are canonically conjugate.
Inserting the full expansion Eq.~(\ref{eq:derivatives}) 
into Eq.~(\ref{eq:first}) yields  
an obvious differential equation in $\psi_x$
with respect to $x$. The canonical transformation   
$i d/dj\rightarrow \tilde v$ and $j\rightarrow -i d/d\tilde v$
preserves the eigenvalues and produces the differential equation 
\begin{equation}
-\frac{\omega^2}2\frac{d^2\psi_{\tilde v}}{d\tilde v^2}-
\left(\sum_{k=0}^\infty\frac{(-1)^k\tilde v^{2k}}{(2k)!}\right)
\psi_{\tilde v}=\lambda\psi_{\tilde v}.
\end{equation}  
Noticing that the sum is equal to $\cos(\tilde v)$ and 
setting $v:=(\tilde v+\pi)/2$, we obtain the canonical form of 
the Mathieu  equation with parameters given in 
Eq.~(\ref{eq:mathparam}).

After these important preliminaries, we are able to proceed towards
the actual solution for the discretised harmonic oscillator.
In order to treat eigenvectors of all matrices $H(x_0)$ on an
equal footing, we replace the index $j$ by $x:=j-x_0$. For
arbitrary values of $x_0$ and $j$ the new index $x$ becomes a continuous
one on $\mathbb{R}$. We thus obtain functions $\psi_x^{(n)}$, where
$n$ is the state index.
We propose that these functions $\psi_x^{(n)}$ are real-analytic and that
they give the eigenvectors of $H(x_0)$ asymptotically, i.e.
\begin{equation}
\psi^{(x_0)}_n\sim\{\psi^{(n)}_{j-x_0}\}_{j=-\infty}^\infty
\end{equation}
when $\omega\rightarrow 0$. The problem tends to the continuous 
one in the lowest (infinitesimal) approximation in $\omega$. Thus it is
reasonable to assume that the lowest-order approximation for the 
solution functions is given by $\psi_x^{(n)}\sim \psi_n(x)$
as $\omega\rightarrow 0$.

The general form of the asymptotical solution of the discretised
harmonic oscillator now reads  
\begin{equation}
\psi_x^{(n)}\propto \exp\left(\sum_{k=1}^\infty\sum_{l=k}^\infty
\alpha_{kl}^{(n)}\omega^{l-1}\xi^{2k}\right)
\sum_{k=0}^{k'}\sum_{l=1}^\infty \left(h^{(n)}_k\omega^{l-1}
\beta^{(n)}_{kl}\xi^{n+2(k-k')}\right),
\label{eq:genansatz}
\end{equation}
where $\alpha_{kl}^{(n)}$ and $\beta_{kl}^{(n)}$ are constants 
to be determined. The solution to the continuous case yields
$\alpha_{1,1}^{(n)}=-1/2$ and $\beta_{k,1}^{(n)}=1$. We are free
to normalise the solution so we can choose
$\beta_{0,l}^{(n)}=0$ for $l>1$. 

The main point of introducing the functions $\psi_x^{(n)}$ is
that they transform the difference-equation-type 
eigenvalue problem corresponding to
the discretised harmonic oscillator into an infinite set
of algebraic equations for each value of $n$. 
The eigenvalues~(\ref{eq:eigenexp}) appear as parameters
and they are required in order to solve the equations for
the sets of coefficients $\{\alpha_{kl}^{(n)}\}$ and 
$\{\beta_{kl}^{(n)}\}$. Fortunately, the equations uniquely
determine every single coefficient. Because the expansion
of the eigenvalues is asymptotical, the meaning of the
full solution to these equations must be determined later.

In practice, we need a suitable truncation of 
Eq.~(\ref{eq:genansatz}) and thus we define an 
(unnormalised) approximate eigenvector
\begin{equation}
\psi_n^{(m,x_0)}:=\{\psi_{j-x_0}^{(n,m)}\}_{j=-\infty}^{\infty},
\end{equation}
where $\psi_{x}^{(n,m)}$ contains only those terms with $l\le m$.
The definition of $\psi_n^{(1,x_0)}$ obviously coincides with
the continuous solution at $x_0$. In numerical calculations, 
and always for even values of $m$, we must truncate the eigenvector
with respect to $j$, by setting $(\psi_n^{(m,x_0)})_j=0$ for
components $\vert j\vert >j_0$ with a sufficiently large value of 
$j_0$.

We now give the infinite set of algebraic
equations corresponding to the transformation of the
difference equation when the solution functions 
$\psi_x^{(n)}$ are substituted into the eigenvalue
equation.  Rearranging the terms, we find
that each equation can be written in the form 
\begin{equation}
\frac{\psi_{x-1}^{(n)}+\psi_{x+1}^{(n)}}{2}=
\psi_x^{(n)}(-\lambda_n+\omega^2x^2/2),\label{eq:general}
\end{equation}
where $x=j-x_0$. Inserting the general ansatz~(\ref{eq:genansatz})
into Eq.~(\ref{eq:general}) expresses the equation in terms
of $\xi$ and $\omega$. 
The exponential part of the ansatz 
on the right-hand-side  canceled simply by substracting the
corresponding exponent from those on the left-hand-side.
This yields an equation
\begin{eqnarray}
&&\frac12\left[\exp\left(\sum_{k=1}^\infty\sum_{l=k}^\infty
\alpha_{kl}^{(n)}\omega^{k+l-1}[(x-1)^{2k}-x^{2k}]\right)
\sum_{k=0}^{k'}\sum_{l=1}^\infty \left(h^{(n)}_k\omega^{l-1}
\beta^{(n)}_{kl}[\sqrt{\omega}(x-1)]^{n+2(k-k')}\right)+\right.\cr
&&\ \left.\exp\left(\sum_{k=1}^\infty\sum_{l=k}^\infty
\alpha_{kl}^{(n)}\omega^{k+l-1}[(x+1)^{2k}-x^{2k}]\right)
\sum_{k=0}^{k'}\sum_{l=1}^\infty \left(h^{(n)}_k\omega^{l-1}
\beta^{(n)}_{kl}[\sqrt{\omega}(x+1)]^{n+2(k-k')}\right)\right]=\cr
&&
\left[-\sum_{m=0}^\infty \lambda_{n(m)}\,\omega^m+\frac{\omega^2x^2}2
\right]\sum_{k=0}^{k'}\sum_{l=1}^\infty \left(h^{(n)}_k\omega^{l-1}
\beta^{(n)}_{kl}[\sqrt{\omega}\,x]^{n+2(k-k')}\right).\label{eq:expanded}
\end{eqnarray}
These equations are then expanded as functions of $x$ and $\omega$ as 
the resulting equations are easier to solve.  The equations must 
hold for all values
of the linearly independent variables $x$ and $\omega$ 
so each  equation must be solved separately. For the purposes of 
generality, it would be preferable to expand with respect to $\xi$, 
but the resulting equations are much more difficult, both to obtain
and to solve. Nevertheless, the obtained solution can be inserted
into to these equations in order to show that
the results are correct. This will be done in Sec.~\ref{sec:proof}.

In order to obtain the eigenvector $\psi_{n}^{(m,x_0)}$ we must 
solve and satisfy all equations corresponding to 
\begin{equation}
\{\{\omega^{m'}\xi^{n+2m'-2l'}\}_{l'=0}^{m'+k'}\}_{m'=0}^m.
\label{eq:firstord}
\end{equation}
This is of course done recursively, by inserting the known part
of the solution and solving for the next level. In order to 
connect Eq.~(\ref{eq:firstord}) with the order of the solution,
we state that the equations corresponding to a fixed value of
$m'$ uniquely determines the coefficients with $l=m'$. 

After a while, one starts to see regularities in the coefficients
and attempts to express these in a functional form. We have been
able to find rather general expressions for the coefficients.
This means that the coefficients have been expressed in terms 
of $n$ and the order of the expansion, whenever this is possible.
We have conjectured the general form of the terms which means
that we know how far away we are from obtaining further terms.

We have found that the functions $\psi^{(n,m)}_x(\xi)$ are asymptotical
solutions of the differential equation
\begin{equation}
\left(-\sum_{k=0}^m\frac{\omega^k}{(2k)!}\frac{d^{2k}}{d\xi^{2k}}+
\frac{\omega \xi^2}{2}\right)\psi^{(n,m)}_x=\sum_{k=0}^m\lambda_n^{(k)}
\omega^k\psi^{(n,m)}_x\label{eq:difftrun}
\end{equation}
in a specific sense. After all derivatives have been carried out, 
the terms multiplying the common exponential factor cancel  
up to and including the order $\omega^m$. If the solution 
$\psi^{(n,m-1)}_x$ is known, we obtain an explicit differential
equation for the exponential part $\omega^{m-1}f_m(\xi)$, the
correction to the Hermite polynomial $\omega^{m-1}g_m(\xi)$ and
the energy eigenvalue $\lambda_n^{(m)}$. In case of the ground
state and the first excited state ($n=1$), the condition $g_m(\xi)=0$
renders the problem solvable. For $n\ge 2$ we must insert the
ansatz~(\ref{eq:genansatz}) in order to obtain
the solution. 

The results are, naturally, in complete agreement with those 
obtained by using the difference equation. We are using just
another representation of the original problem. 
Equation~(\ref{eq:difftrun}) enables us to obtain the solutions
for fixed values of $n$
up to relatively high orders with respect to powers of $\omega$.
Thus we can both extend the general expression for the
eigenenergies in Eq.~(\ref{eq:eigenexp}) and those for 
the coefficients in the
exponential part of the solutions. For the ground state energy, 
we find the terms beyond order $\omega^{16}$ to be 
\begin{eqnarray}
&&-\frac{363372562420411197\,\omega^{17}}{2^{79}}
-\frac{6258692522467212813\,\omega^{18}}{2^{83}}-
\frac{227867608383920243815\,\omega^{19}}{2^{88}}-\cr
&&\ \ \frac{4372199488222446620121\,\omega^{20}}{2^{92}}-
\frac{352807992522448740907163\,\omega^{21}}{2^{98}}-
  \frac{7465886451386334274097895\,\omega^{22}}{2^{102}}-\cr
&&\ \  \frac{330752735437897260202410959\,\omega^{23}}{2^{107}}-
  \frac{7654237307570898665851927581\,\omega^{24}}{2^{111}}-\cr
&&\ \  \frac{1477812451863756884805687589129\,\omega^{25}}{2^{118}}-
  \frac{37132718819258763418452357390369\,\omega^{26}}{2^{122}}-\cr
&&\ \  \frac{1939848955425261040700592191917783\,\omega^{27}}{2^{128}}-
  \frac{52598573101029275526869814635336865\,\omega^{28}}{2^{131}}-\cr
&&\ \  \frac{5914101566562517015636997146651378649\,\omega^{29}}{2^{137}}-
  \frac{172129355454985486683952198830698506149\,\omega^{30}}{2^{141}}-\cr
&&\ \  \frac{10362392343003738344189045786484697182753\,\omega^{31}}{2^{146}}
  +\mathcal{O}(\omega^{32}).
\end{eqnarray}
The corresponding asymptotical eigenvector contains $31(1+31)/2=496$ 
linearly independent terms. The coefficient of $\omega^{30}x^2$ in
the exponential part reads
\begin{equation}
-\frac{5207328980459439428858189871778019425519567564728193}
{2765292404617797269550429065808396826741571584}.
\end{equation}
The general solution is given in the next section.

\section{The general solution of the discretised harmonic oscillator
\label{sec:solu}}

For reasons of completeness and easy accessibility some
of the definitions will be repeated in this section.
The $m^{\mathrm{th}}$ order solution function 
$\psi_x^{(n,m)}$,  corresponds all terms up to and including 
$l=m$ in Eq.~(\ref{eq:genansatz}). The asymptotical expansion of 
the eigenvalues $\lambda_n$ is  given in Eq.~(\ref{eq:eigenexp}).
The state index $n$ determines
two expansion parameters, 
\begin{equation}
\hat n:=2n+1\quad\mathrm{and}\quad k':=\lfloor n/2\rfloor,
\end{equation}
where $k'=n/2$ if $n$ is even and $k'=(n-1)/2$\ if $n$ is odd.
The gamma function $\Gamma(x)$ is the generalised factorial
with the defining property $x\Gamma(x)=\Gamma(x+1)$. We 
need the values for integer and half-integer values which read
\begin{equation}
\Gamma(k)=(k-1)!,\quad \Gamma(k+\mbox{$\frac12$})=2^{-k}
\sqrt\pi (2k-1)!!,
\end{equation}
where the double factorial $k!!$ for integer values
of $k$ is given by $k(k-2)\times\cdots \times (1\ \mathrm{or}\ 2)$. 
The coefficients in the Hermite polynomials simplify to
\begin{equation}
h^{(n)}_k=\frac{(-1)^{k'+k}2^{2k+(1-(-1)^n)/2} n!}
{(2k+(1-(-1)^n)/2)!(k'-k)!}.
\end{equation}
A convenient normalisation for the eigenvectors is obtained 
by requiring that
\begin{equation}
\psi^{(n)}_x \sim \xi^{(1-(-1)^n)/2)},\quad x\rightarrow 0,
\end{equation}
i.e. $\sim 1$ for even values of $n$ and $\sim \xi$ for 
odd values of  $n$. Also bear in mind that
\begin{equation}
\alpha_{1,1}^{(n)}=-1/2,\quad \beta_{k,1}^{(n)}=1,\quad 
\mathrm{and}\quad \beta_{0,l(>1)}^{(n)}=0. 
\end{equation}

Under these constraints we have conjectured that the general form 
of the coefficients. In the exponential part,
\begin{equation}
\exp\left(\sum_{k=1}^\infty\sum_{l=k}^\infty
\alpha_{kl}^{(n)}\omega^{l-1}\xi^{2k}\right),\label{eq:exppart}
\end{equation}
the coefficients can be written as
\begin{equation}
\alpha_{k,k+l}^{(n)}=\sum_{l'=0}^l\alpha_{k,k+l}^{[l']}\hat n^{l'}.
\label{eq:genalpha}
\end{equation} 
Please note that if the coefficient 
$\alpha_{kl}^{(n)}$ are written as polynomials in
$n$ instead of $\hat n$, the signs of the corresponding 
expansion coefficients
$\tilde\alpha_{k,k+l}^{[l']}$ appear to be given by  $(-1)^k$. 
An efficient way
to write these coefficients is given by 
\begin{equation}
\alpha_{k,k+l}^{(n)}=(-1)^k 2^{-2k}\left(\sum_{l'=0}^{\lfloor l/2\rfloor}
\frac{\Gamma(k+1/2)Q(k,l,2k-5+l')\hat n^{l-2l'}}{\Gamma(k+l)\sqrt{\pi}}
+\sum_{l'=0}^{\lfloor (l-1)/2\rfloor}\bar Q(k,l,k-2+l')\hat n^{l-1-2l'},
\right)\label{eq:realgenalp}
\end{equation}
where $Q(k,l,2k-5+l')$ and $\bar Q(k,l,k-2+l')$ are polynomials in
$k$ of orders $2k-5+l'$ and $k-2+l'$, respectively. 
An important consequence of Eq.~(\ref{eq:realgenalp}) is that
regardless of the values of  $l$ and $n$ we have 
\begin{equation}
\lim_{k\rightarrow\infty}\alpha_{k+1,k+1+l}^{(n)}/\alpha_{k,k+l}^{(n)}=-1/4.
\end{equation}

The explicit expressions for the seven leading coefficients 
have been obtained and they read: 
\begin{eqnarray}
\alpha_{kk}^{(n)}&=&\frac{(-1)^k 2^{2-2k}\Gamma(k+1/2)}{
k(2k-1)^2\Gamma(k)\sqrt\pi},\cr
\alpha_{k,k+1}^{(n)}&=&(-1)^k 2^{-2-2k}\left(\frac1k+\frac{\Gamma(k+1/2)}
{\Gamma(k+1)\sqrt\pi}\frac{\hat n}{k}\right),\cr
\alpha_{k,k+2}^{(n)}&=&(-1)^k 2^{-4-2k}\left(\hat n+\frac{\Gamma(k+1/2)}
{24\Gamma(k+2)\sqrt\pi}\left[(3 + 52k + 40k^2)+(9+12k)\hat n^2\right]
\right),\cr
\alpha_{k,k+3}^{(n)}&=&(-1)^k 2^{-9-2k}\left\{
 (-1 + 7k + 5k^2) + (3 + 4k)\hat n^2
+\frac{\Gamma(k+1/2)}
{24\Gamma(k+3)\sqrt\pi}\times\right.\cr
 &&\ \,\left. \left[(243 + 1119k + 1928k^2 + 1376 k^3 + 320k^4)\hat n+
   (33 + 101k + 104k^2 + 32k^3)\hat n^3\right]\right\},\cr
\alpha_{k,k+4}^{(n)}&=&(-1)^k 2^{-14-2k}\left\{
(53 + 120k + 136k^2 + 40k^3)\hat n + (37 + 72k + 32k^2)\hat n^3/3
+\frac{\Gamma(k+1/2)}
{48\Gamma(k+4)\sqrt\pi}\times\right.\cr
 &&\ \,\left[(-2612925 - 5292132k + 10675063k^2 + 
  36766856k^3 + 40148416k^4 + 21300608k^5 + 
  5544448k^6 +\right.\cr
  &&\ \ \ 565760k^7)/315+
   (11070 + 60044k + 130810k^2 + 142112k^3 + 81280k^4 + 23168k^5 + 2560k^6)
   \hat n^2+\cr
  &&\ \ \left.\left.(585 + 2288k + 3585k^2 + 2696k^3 + 960k^4 + 128k^5)
   \hat n^4\right]\right\},\cr
\alpha_{k,k+5}^{(n)}&=&(-1)^k 2^{-20-2k}\left\{
  (-5187 - 672k + 6580k^2 
  + 7684k^3 + 3164k^4 + 452k^5)/3+(1214 + 3744 k +\right.\cr
 &&\ \  4080 k^2 + 1968 k^3 + 320 k^4)\hat n^2+
   (345 + 808k + 576k^2 + 128k^3)\hat n^4/3+
    \frac{\Gamma(k+1/2)}{48\Gamma(k+5)\sqrt\pi}\times\cr
 &&\ \ \left[ 
   (740893230 + 3944788389 k + 9627147810 k^2 + 14943869467 k^3 +
     15287941200 k^4 +\right.\cr
 &&\ \ \, 10116675072 k^5 + 4238798592 k^6 + 
    1079918592 k^7 + 152076288 k^8 + 9052160 k^9)\hat n/315+\cr
 &&\ \ (1825740 + 11037114 k + 27955236 k^2 + 37919062 k^3 + 
    30169312 k^4 + 14491648 k^5 + 4122880 k^6 +\cr
 &&\ \ \, 636928 k^7 + 40960 k^8)\hat n^3/3+(85050 + 
    381087 k + 729798 k^2 + 752369 k^3 + 447024 k^4 +\cr
 &&\ \ \,\left.\left.	 
    152576 k^5 + 27648 k^6 + 2048 k^7)\hat n^5/5\right]\right\},\cr
\alpha_{k,k+6}^{(n)}&=&(-1)^k 2^{-26-2k}\left\{(
  378033 + 496368k + 786528k^2 + 710816k^3 + 339904k^4 + 
   79552k^5 + 7232k^6)\hat n/3+\right.\cr
 &&\ \ (69714 + 241312k + 303392k^2 + 177696k^3 + 49408k^4 + 5120k^5)
    \hat n^3/3+\cr
 &&\ \ (17217 + 45360k + 40960k^2 + 15360k^3 + 2048k^4)
	\hat n^5/15+\frac{\Gamma(k+1/2)}
	{180\Gamma(k+6)\sqrt\pi}\times\cr
 &&\ \,\left[(-24640192386810 - 105728184475128k - 155775948330744k^2 - 
   74654535511116k^3 +\right.\cr 
 &&\ \ \,74660144680858k^4 + 156803802177352k^5 + 134434233033760k^6 + 
  70722102090816k^7 +\cr 
 &&\ \ \, 24590691451392k^8 + 5680345583616k^9 + 
  839668527104k^{10} + 71921254400k^{11} +\cr 
 &&\ \ \, 2714009600k^{12})/9009+(22093103970 + 162201234402k + 
   504160865145k^2 + \cr 
 &&\ \ \, 882850470198k^3 + 986932878421k^4 + 745434338828k^5 + 
    388089936864k^6 + \cr 
 &&\ \ \, 138972684672k^7 + 
    33504543744k^8 + 5179637760k^9 + 462565376k^{10} + 18104320k^{11})
   \hat n^2/21+\cr
 &&\ \ (152041050 + 991922940 k + 2784482730 k^2 + 4353707520 k^3 + 
    4203836660 k^4 +\cr
 &&\ \ \, 2632731680 k^5 + 1088777440 k^6 + 
    294912320 k^7 + 50245120 k^8 + 4874240 k^9 + 204800 k^{10})\hat n^4+\cr
 &&\ \ (2606310 + 12799746k + 27798345k^2 + 34245070k^3 + 26181505k^4 +\cr 
 &&\ \ \, \left. \left .12857468k^5 + 
  4055200k^6 + 792320k^7 + 87040k^8 + 4096k^9)\hat n^6\right]\right\}
   .\nonumber
\end{eqnarray}
Thus, for an arbitrary order $m$, we can obtain the expressions for 
coefficients corresponding to $\{\omega^{m-1}\xi^{2m-l'}\}_{l'=0}^6$.
Furthermore we find
\begin{eqnarray}
\alpha^{(n)}_{1,8}&=& -(505549159\hat n + 177209155\hat n^3 +
      8289645\hat n^5 +  40329\hat n^7)/2^{37} - \cr
	  &&\ \ (-2741702 + 12248825\hat n^2 + 1518052\hat n^4 + 
	  26073\hat n^6)/2^{32},\cr
\alpha^{(n)}_{1,9}&=& -(-840819020949 + 1419128841068\hat n^2 + 
    221074444682\hat n^4 + 6195597884\hat n^6 + 
          21259875\hat n^8)/(3\times 2^{47}) - \cr
   &&\ \ (1318785849\hat n + 459389255\hat n^3 + 
        29718111\hat n^5 + 335617\hat n^7)/2^{38},\cr     
\alpha^{(n)}_{2,9}&=& 	(131257276187\hat n + 37843099187\hat n^3 + 
     1323046497\hat n^5 +  4456305\hat n^7)/2^{44} +\cr
	&&\ (48228434 + 93959845\hat n^2 + 8787700\hat n^4 +
        110661\hat n^6)/2^{34}.\nonumber
\end{eqnarray}
The exponential part~(\ref{eq:exppart}) is now completely determined up
to the ninth order, i.e. known for arbitrary values of $n$ for terms
with $l\le 9$. 

If we exclude the dependence on $2^{-2k}$ and also that given by 
the gamma functions in the coefficients, we observe a very
distinct regularity. The dependence of the leading power $k$ in
each polynomial sequence starting from $\hat n^0$ for a fixed value 
$l$ in $\alpha_{k,k+l}^{(n)}$ and going upwards by one both for $l$
and the power of $\hat n$ is so far always given by   
\begin{equation}
\{\tilde\alpha_{(l,l')}\}_{l'}=\{\tilde\alpha_{(l,0)}/(4^{l'}(l')!)\}.
\end{equation}
The initial values in cases $l'\le 5$ are given by
\begin{equation}
\{\tilde\alpha_{(l,0)}\}_{l=0}^5=\{1,1/4,5/48,5/512,221/96768,113/786432\}.
\end{equation}
This dependence is by no means proven but it corroborates our choice
for the prefactors in Eq.~(\ref{eq:realgenalp}).

The coefficients $\{\beta^{(n)}_{kl}\}$ determine a set of
new polynomials,
where the coefficients multiplying the powers of $\xi$ depend on 
$\omega$. In the limit $\omega\rightarrow 0$ 
these polynomials tend to the Hermite polynomials. They are,
unquestionably, a new class of generalised Hermite polynomials.
They are defined as parts of the eigenvectors of a Hermitian
matrix. Because the exponential part is rather complicated,
the measure, with respect to which they become asymptotically
orthogonal, is necessarily a complicated one.
It depends on both of the eigenvectors, i.e., it is not 
a measure in the classical sense at all. No simple 
recursion relation for the polynomials is yet known, 
and we do not know, whether they satisfy any differential
equation of finite order. This means that they could be an
example of the second category of generalised Hermite
polynomials as defined by Borzov in Ref.~\onlinecite{bor01}.
Such discussion is beyond the scope of the present study
and we will concentrate on simpler properties of the polynomials.

Our generalised Hermite polynomials are defined as
\begin{equation}
H_n^{\omega}(\xi):=\sum_{k=0}^{k'}\tilde h^{(n)}_k\xi^{n+2(k-k')},
\label{eq:genhermitepol}
\end{equation}
where the modified coefficients are given by
\begin{equation}
\tilde h^{(n)}_k:=h^{(n)}_k \sum_{l=1}^\infty \left(\omega^{l-1}
\beta^{(n)}_{kl}\right).\label{eq:modcoeff}
\end{equation}
Because the generalised  Hermite polynomials $H_n^{\omega}(\xi)$ 
fix the nodes (zeroes) of the functions $\psi^{(n)}_x$, it
is equally important to obtain correct polynomials as it
is to obtain the correct exponential factors. 

We conjecture that the general form of the coefficients 
$\{\beta^{(n)}_{kl}\}$ reads
\begin{equation}
\beta^{(n)}_{kl}=\sum_{l'=0}^{l-1}\left(\sum_{\bar l=1}^{2(l-1)
-l'}(\rho^{(l)}_{l'\bar l}+[(1-(-1)^n)/2]
\bar\rho^{(l)}_{l'\bar l})
k^{\bar l}\right)(k')^{l'},\label{eq:genbeta}
\end{equation}
where $\rho^{(l)}_{l'\bar l}$ and $\bar\rho^{(l)}_{l'\bar l}$ 
are constants.  Additionally, $\bar\rho^{(l)}_{l'\bar l}=0$ 
when $\bar l=2(l-1)-l'$ or $l'=l-1$. 
This expansion with respect to $k$ and $k'$ shows that 
even and odd values of $n$ should be treated separately. 

Some general properties of the coefficients $\rho^{(l)}_{l'\bar l}$
and $\bar\rho^{(l)}_{l'\bar l}$ have been gleaned. The recurring 
appearance of the factor $(10k'-k)$ is by far the most striking of 
the observed regularities. This factor may, in time, explain some
properties generalised Hermite polynomials. We conjecture that
\begin{equation}
\sum_{l'=0}^{l-1}\left(\rho^{(l)}_{l',2(l-1)-m_0-l'}k^{2(l-1)-m_0-l'}
n^{l'}\right)
=k^{l-1-m_0}(10k' - k)^{l-1-2m_0}P(2m_0,l),\label{eq:kcommak}
\end{equation}
where  $2m_0<l$ and $P(2m_0,l)$ denotes a $(2m_0)^{\mathrm{th}}$ order
polynomial in $k$ and $k'$. 
Similarly, the difference between even and odd values of $n$
corresponds to 
\begin{equation}
\sum_{l'=0}^{l-2}\left(\bar\rho^{(l)}_{l',2l-3-m_0-l'}k^{2l-3-m_0-l'}
n^{l'}\right)
=k^{l-1-m_0}(10k' - k)^{l-2-2m_0}\bar P(2m_0,l),\label{eq:kcommak2}
\end{equation}
where $2m_0<l-1$ and $\bar P(2m_0,l)$ again denotes 
a $(2m_0)^{\mathrm{th}}$ order polynomial in $k$ and $k'$.

In the leading and next-to-leading orders the polynomials 
$P(0,l)$, $P(2,l)$, $\bar P(0,l)$, and $\bar P(2,l)$ have been
explicitly evaluated.
Thus, we define the quantities
\begin{equation}
B(l,2l-2):=
\frac{k^{l-1}(10k' - k)^{l-1}}{48^{l-1}(l-1)!},\label{eq:betalead1}
\end{equation}
and
\begin{equation}
B(l,2l-3):=
\frac{k^{l-2}(10k' - k)^{l-3}}{5\times 48^{l-1}(l-2)!}
[(l-2)658(k')^2+(402-126l)k'k+(8l-31)k^2],\label{eq:betalead2}
\end{equation}
which has been confirmed up to the sixth order, i.e. $l=6$. Similarly, the 
leading differences give rise to the quantities 
\begin{equation}
\bar B(l,2l-3):=
\frac{4k^{l-1}(10k' - k)^{l-2}}{48^{l-2}(l-2)!}\label{eq:betadiff1}
\end{equation}
and
\begin{equation}
\bar B(l,2l-4):=
\frac{k^{l-2}(10k' - k)^{l-4}}{5\times 48^{l-1}(l-3)!}
[(2632l-2576)(k')^2+(1470-504l)k'k+(32l-145)k^2].\label{eq:betadiff2}
\end{equation}
The leading terms are very similar, but also the next-to-leading
terms $B(l,2l-3)$ and  $\bar B(l,2l-4)$ share several common features.
Most importantly, the $l$-dependence in the polynomial section
is identical, apart from a factor of $4$.  

Below, we give the  explicit values of the 
coefficients $\beta^{(n)}_{kl}$ in cases  $2\le l\le 7$.
It is convenient to separate the even and odd values of $n$,
because the correct expansion parameter appears to be $k'$.
Please note that these expressions automatically yield
$\beta^{(n)}_{0,l}=0$ for $l>1$.
\begin{eqnarray}
\beta^{(2k')}_{k,2}&=&\left[(3k-k^2)+(10k)k'\right]/48,\cr
\beta^{(2k'+1)}_{k,2}&=&\beta^{(2k')}_{k,2}+k/12,\cr
\beta^{(2k')}_{k,3}&=&\left[(855k-64k^2-14k^3+5k^4)+(784k+48k^2-100k^3)k'
+(1316k+500k^2)(k')^2\right]/23040,\cr
\beta^{(2k'+1)}_{k,3}&=&\beta^{(2k')}_{k,3}+
\left[(249k+49k^2-20k^3)+(532k+200k^2)k'\right]/11520,\cr
\beta^{(2k')}_{k,4}&=&B(4,6)+B(4,5)+
  \left[(371385k-203498k^2-12129k^3+1438k^4)+\right.\cr
    &&\ (1110698k+102042k^2-26252k^3)k'+
     \left.(496932k+93984k^2)(k')^2+
       560200k (k')^3\right]/23224320,\cr
\beta^{(2k'+1)}_{k,4}&=& \beta^{(2k')}_{k,4}
  + \bar B(4,5) + \bar B(4,4)+ 
  \left[(67680k+12347k^2-2602k^3)+\right.\cr 
&&\ \left.(108544k+33762k^2)k' + 
   114456k (k')^2\right]/3870720.\cr
\beta^{(2k')}_{k,5}&=&B(5,8)+B(5,7)+
  \left[(278751375k-202014918k^2+35222268k^3+4026748k^4
+28158k^5-9944k^6)+\right.\cr
    &&\ (713250468k-281790420k^2-61452368k^3-196176k^4+
	  209856k^5)k'+\cr
   &&\ (1105743252k+198178852k^2-10630680k^3-275344k^4)(k')^2+\cr
   &&\ (319197168k+81282336k^2-22799744k^3)(k')^3
     +\left.(271672512k+148408976k^2)(k')^4    
      \right]/22295347200,\cr
\beta^{(2k'+1)}_{k,5}&=&\beta^{(2k')}_{k,5}+\bar B(5,7)+\bar B(5,6)+
     \left[(119817225k-23468037k^2-12060122k^3-
     330312k^4+100198k^5)+\right.\cr
     &&\ (436319556k+103769756k^2-5886336k^3-1921112k^4)k'+\cr
     &&\ (321608148k+118383396k^2-904080k^3)(k')^2+
	 \left.(224147856k+120486304k^2)(k')^3\right]
     /11147673600\cr
\beta^{(2k')}_{k,6}&=&B(6,10)+B(6,9)+\left[(134035780725k - 166751340588k^2 
  +39327194883k^3- 2269605874k^4-\right.\cr
  &&\ 477614210k^5 - 19226552k^6 +221782k^7 + 484k^8)+(413990823078k - 
  217584747090k^2+\cr
  &&\ + 22678956764k^3 + 8841166604k^4 + 479019924k^5 - 6549884k^6 + 
        566984k^7)k'+\cr 
  &&\ (526339688532k\hskip-1pt-\hskip-1pt 155591533528k^2\hskip-1pt - 
   \hskip-1pt 55003198072k^3 \hskip-1pt-\hskip-1pt 
    3518713436k^4 \hskip-1pt+\hskip-1pt 85514000k^5\hskip-1pt - 
	 \hskip-1pt 24903296k^6)(k')^2+\cr
  &&\ (556945898088k + 131085561976k^2 + 2790556248k^3 - 
    280060176k^4 + 424040144k^5)(k')^3+\cr
  &&\ (116760015552k + 34523271136k^2 - 5865150192k^3 - 
    3338174576k^4)(k')^4+\cr
   &&\left. (79966766400k + 46102886720k^2 + 10162787360k^3)(k')^5
      \right]/11771943321600,\cr
\beta^{(2k'+1)}_{k,6}&=&\beta^{(2k')}_{k,6}+\bar B(6,9)+\bar B(6,8)+
     \left[(34460588160k - 26910050283k^2 + 72069996k^3 + 1282383895k^4
	  +\right.\cr 
  &&\ 103465570k^5 - 948002k^6  -  316976k^7) + (216801198648k  -  
   21671791146k^2  -  \cr  
  &&\ 18471533106k^3- 1699322576k^4+78651584k^5 + 5865684k^6)k'+ 
     \cr 
  &&\ (345295895928k + 96181762100k^2 +1478206984k^3 -1409258180k^4 + 
    60822872k^5)(k')^2+\cr
  &&\ (160052617776k + 64547633160k^2 + 3204992824k^3 - 
    1898232512k^4)(k')^3+\cr
&&\ \left.(83156900448k + 47130830560k^2 + 10276562912k^3
    )(k')^4\right]/5885971660800,\cr
\beta^{(2k')}_{k,7}&=&B(7,12)+B(7,11)+\left[(21167446950775125k - 
34318046368345140k^2 +13674300462898392k^3 -  \right.\cr
&&\ 1352901404372446k^4 -2843855572731k^5 + 11311875159790k^6 + 
 704407032828\ k^7 +\cr
&&\ 12949326156k^8 - 177366189k^9 + 37677640k^{10})+
    (59570630372492640 k -\cr
&&\ 60644270495554704k^2 + 10066261151648252k^3 + 
    104602336760652k^4 - 246415137367020k^5-\cr
&&\ 20207362771548k^6 - 460168946016k^7 + 2604105504k^8 - 
     2483040560k^9)k'+\cr
&&\ (99669485611466412k - 39020273844707836k^2 + 2200698814542984k^3 
    + 2070713072954600k^4+\cr
&&\ 212428368788100 k^5 + 5622413614220 k^6 + 
    82814211480 k^7 + 70784553840 k^8)(k')^2+\cr
&&\ (82488078028378080 k - 18442822328400480 k^2 - 9100818756007520
    k^3 - 964844434165920 k^4 -\cr 	
&&\ 24156442527360k^5 - 3013682511840k^6 - 1116228072960k^7)(k')^3+
   (66588038149135200 k +\cr
&&\ 18345507366303440k^2 + 1191268975557840k^3 + 
    10518809509520k^4 + 42474114642960k^5 +\cr
&&\ 10244315921840k^6)(k')^4+(10839030004200960k + 
     3516288982521792k^2 - 363895953410496k^3-\cr
&&\  304221200739456k^4 - 51610667908800k^5)(k')^5+
    (6218212960526208k + 3705496740373376k^2+\cr
&&\left.\ 898601964676416k^3 + 110684037464000k^4)(k')^6
    \right]/1542595452862464000,\cr
\beta^{(2k'+1)}_{k,7}&=&\beta^{(2k')}_{k,7}+\bar B(7,11)+\bar B(7,10)+
     \left[(-460686821541975k - 1941941074537755k^2 
	   + 366877584331212k^3+
	  \right.\cr 
  &&\ 41494582964306k^4 - 9965970910165k^5 - 1112954021925k^6 - 
    34972438722k^7+ 958101144k^8+\cr
  &&\  3798795k^9)+(9145976126266080 k -
      3823250702059872k^2-191143081971676k^3 + \cr
  &&\  173415983605096k^4 +24948282593576k^5 + 
    815890000796k^6 -41300603344k^7 + 1974092120k^8 )k'+\cr
  &&\ 
     (19045045703842332k -607795420829248k^2-1422456568355676k^3 - 
	  195873935369616k^4 -\cr
  &&\ 2655255816252k^5 +611948653316 k^6 -98001423520k^7 )(k')^2+(
       18941405236672032 k+\cr
  &&\ 5863460843089248k^2 + 345511721120640k^3 - 55753844615456k^4 - 
      654723253184k^5 +\cr
  &&\ 1792716525600k^6 )(k')^3+
      (6296099301099168k + 2692631923242160k^2 +232744611197904k^3 -\cr
  &&\  59049744898144k^4 - 14606140268720k^5)(k')^4+
       (2605202959125888k + 1525593370591680k^2 +\cr
  &&\ \left. 365590616623232k^3 + 
      44670372947200k^4)(k')^5\right]/257099242143744000.\nonumber
\end{eqnarray}
In combination with the exponential parts these 
coefficients determine explicit, analytical 
expressions for solution function $\psi_x^{(n,7)}$ for
arbitrary values of $n$.\cite{matlab}

It must be re-emphasised that Eq.~(\ref{eq:genansatz}) is an 
asymptotical solution. Two partially overlapping reasons 
for this behaviour must be stated. First, the solution
depends on two independent length scales, i.e. $x$ and $\omega$,
and second, the coordinate transformation $x\mapsto \xi=
\sqrt{\omega}x$ is singular at $\omega=0$. These points are
rather extensively covered in Ref.~\onlinecite{boy99}.
The eigenvalues are asymptotically
exact for even values of $n$ at $x_0=0$  and, probably, 
for odd values of
$n$ at $x_0=\pm\mbox{$\frac12$}$. Because the error decays 
exponentially in $1/\omega$, this dependence on $x_0$
vanishes much before the asymptotical behaviour
of Eq.~(\ref{eq:asymp}), i.e 
$\Vert \psi_n^{(m,x_0)}-\psi_n^{(x_0)}\Vert\sim C(n,m)\omega^m$,
appears.  

Comparison against numerically obtained eigenstates allows us
to give an approximate expression for the function $C(n,m)$.    
The  validity of the calculations is limited by the numerical precision, 
i.e. to norms of the order of $10^{-11}$--$10^{-12}$. We have employed 
the reliable diagonalisation routines of \textsc{Matlab} 
software for this purpose. We have studied 
eigenvectors up to $n\approx 40$--$50$ and the corresponding 
asymptotical solutions $\psi_n^{(m,x_0)}$ up to the fifth
order. A reasonable, order-of-magnitude estimate for the 
error in the  Euclidean norm, when $n\le 40$, is given by
\begin{equation}
C(n,m)\approx c_m\hat n^{2m},\label{eq:asympacc} 
\end{equation}
where
\begin{equation}
c_1=0.03,\ \ c_2=0.002,\ \ c_3=0.0006,\ \ c_4=1.5\times 10^{-6},
\ \ \mathrm{and}\ \ c_5=3\times 10^{-8}.
\end{equation}
There is a slight difference between even and odd cases, but this
is insignificant in an estimate like this.  The value  of $c_5$ 
is set to fit the observed trend in the other coefficients
as the asymptotical behaviour is only glimpsed. 
In cases $m=2$ and $m=4$, it is vitally important
to remember to truncate the asymptotical eigenvector 
$\psi_n^{(m,x_0)}$ correctly. 

For larger values of $n$, one needs very small values 
of $\omega$ in order to obtain accurate or even 
reasonable results. But for relatively small values
of $n$, say $n\le 10$, the error is extremely small
at $\omega\approx 0.01$. The strong dependence on 
$n$ means that the first few states can be
obtained to a high precision even for quite strong 
couplings in the neighbourhood of $\omega\approx 0.1$.
We have determined the ground state $n=0$ up to the
$31^{\mathrm{st}}$ order and numerical comparison 
strongly supports the asymptotical behaviour 
$\omega^{m}$ for $m\le 13$.

In order to make the above discussion more concrete,
we explicitly give the second-order solutions as 
functions of $\omega$, $\xi=\sqrt{\omega}\, x$, 
$n$ (not $\hat n$) and $k'$. 
For even values of  $n$ we find the solution function  
\begin{equation}
\psi_x^{(n,2)}=A_{n,x_0}\exp\left(-(\mbox{$\frac12$}+
(3+2n)\omega/32)\xi^2+(\omega/96)\xi^4\right)
\sum_{k=0}^{k'}\left(h_k^{(n)}\xi^{2k}(1+(3k-k^2+10kk')\omega/48)\right),
\end{equation} 
where $A_{n,x_0}$ is a normalisation factor which ensures
that $\Vert \psi_n^{(2,x_0)}\Vert=1$. For odd values of $n$
the result is nearly identical, i.e.
\begin{equation}
\psi_x^{(n,2)}=A_{n,x_0}\exp\left(-(\mbox{$\frac12$}+
(3+2n)\omega/32)\xi^2+(\omega/96)\xi^4\right)
\sum_{k=0}^{k'}\left(h_k^{(n)}\xi^{2k}(1+(7k-k^2+10kk')\omega/48)\right).
\end{equation} 
The tiny difference $3k \rightarrow 7k$ in the generalised Hermite 
polynomial is very important, because
otherwise the asymptotical convergence $\Vert\psi_n^{(m,x_0)}-
\psi_n^{(x_0)}\Vert\sim \omega^2$ does not appear. The 
common exponential part in the 
third order solution function  $\psi_x^{(n,3)}$ reads
\begin{equation}
\exp\left(-(\mbox{$\frac12$}+
(3+2n)\omega/32+(53 + 69n + 21n^2)\omega^2/1536)\xi^2
+(\omega/96+(11+6n)\omega^2/1024)\xi^4-(\omega^2/1280)\xi^6\right).
\end{equation}
The explicit solution function $\psi_x^{(n,m)}$ solves the 
asymptotical eigenvalue equation up to the order $\omega^m$
and yields a normwise convergence of $\sim \omega^m$.

When employing these asymptotical solutions, one should first 
study, how accurate eigenvectors are required for the problem at
hand. The next step is to choose the order of the solution and
the correct truncation with respect to $x$. Then, the 
calculations are performed and the results are obtained,
hopefully faster than with the conventional
approach of numerical diagonalisation.\cite{shir93}

\section{Comments on solving the ansatz\label{sec:ansatz}}

In this section we discuss how to solve the set of 
algebraic equations resulting from Eq.~(\ref{eq:expanded})
as effectively as possible. First we observe that the
zeroth order, i.e. terms proportional to arbitrary powers
of $\xi$ are satisfied by the fact $\exp(0)=1$. 
Next, all equations related to terms
\begin{equation}
\{\omega \xi^{n+2-2l'}\}_{l'=0}^{1+k'}
\end{equation}  
are identically satisfied because of the recursion 
relation~(\ref{eq:hermrecur}) rewritten
in terms of the coefficients $h_k^{(n)}$. A careful reader
notices that terms proportional to $\beta_{k,l=2}^{(n)}$ do
appear, but they identically cancel and thus they are not
constrained in this order. 

From here on, we proceed by recursively solving the coefficients for
the next order and also for sufficiently many values of $n$ so that
all coefficients in the expansions of $\{\alpha_{kl}^{(n)}\}$
and  $\{\beta_{kl}^{(n)}\}$ have been constrained. In reality, we
first obtained the solution function $\psi_x^{(n=0,m=6)}$ and
a poorly formulated expression for arbitrary second-order 
solution, i.e. $\psi_x^{(n,m=2)}$, but let us proceed in the
way this should be done. Because the equation are quite
difficult to handle with pen and paper, we chose
to write and simplify the equations with 
\textsc{Mathematica} software.\cite{mathem}  

We first consider the cases $n=0$ and $n=2$ as simple examples. 
For $n=0$ we expand Eq.~(\ref{eq:expanded}) up to and including 
order $\omega^3$ to find
\begin{eqnarray}
&&\left\{1-\frac{\omega}2+\omega^2
\left(\alpha_{1,2}^{(0)}+
\frac18+\frac{x^2}2\right)+
\omega^3\left[-\frac1{48}+\frac{\alpha_{1,2}^{(0)}}2+\alpha_{1,3}+
\alpha^{(0)}_{2,2}+x^2\left(-\frac14-2\alpha^{(0)}_{1,2}+6
\alpha^{(0)}_{2,2}\right)\right]\right\}=\cr
&&\left[1-\frac{\omega}2+\frac{\omega^2(x^2+1/16)}{2}
+\frac{\omega^3}{512}\right].\label{eq:solving}
\end{eqnarray}
Immediately, we obtain
\begin{equation}
\alpha^{(0)}_{1,2}=-3/32\ \ \mathrm{and}\ \ \alpha^{(0)}_{2,2}=1/96.
\end{equation}
Inserting these into Eq.~(\ref{eq:solving}) gives 
$\alpha^{(0)}_{1,3}=-53/1536$.

In the case $n=2$ and we examine all terms 
below the order of $\omega^4$. 
The generalised Hermite polynomial now reads
\begin{equation}  
H_2^\omega(\xi)=
-2+4\omega x^2(1+\beta_{1,1}^{(2)}\omega+\beta_{1,2}^{(2)}\omega^2
+\beta_{1,3}^{(2)}\omega^3)+\mathcal{O}(\omega^5).
\end{equation}
Expanding all terms and moving them onto the same side yields
the equation
\begin{eqnarray}
0&=&\omega^2\left(\frac{23}{32}+\alpha^{(2)}_{1,2}-2\beta_{1,1}^{(2)}\right)
+\omega^3\left[-\frac{521}{1536}-\frac{5\alpha^{(2)}_{1,2}}{2}+
\alpha^{(2)}_{1,3}+\alpha^{(2)}_{2,2}+\beta_{1,1}^{(2)}-2\beta_{1,2}^{(2)}\right.+\cr
&&\ \ \left.x^2\left(-\frac{43}{16}-12\alpha^{(2)}_{1,2}+
6\alpha^{(2)}_{2,2}\right)\right]+\omega^4
\left[\frac{341}{24576}+\frac{(\alpha^{(2)}_{1,2})^2}{2}
-\frac{5\alpha^{(2)}_{1,3}}{2}+\alpha^{(2)}_{1,4}+\right.\cr
&&\ \ \alpha^{(2)}_{1,2}\left(\frac98-2\beta_{1,1}^{(2)}\right)-
\frac{\beta_{1,1}^{(2)}}4+
\beta_{1,2}^{(2)}-2\beta_{1,3}^{(2)}+
x^2\left(\frac{953}{768}+2(\alpha^{(2)}_{1,2})^2-12\alpha^{(2)}_{1,3}-
37\alpha^{(2)}_{2,2}+\right.\cr
&&\ \ \left.6\alpha^{(2)}_{2,3}+
\left.\alpha^{(2)}_{1,2}\left(
\frac{29}2-10\beta_{1,1}^{(2)}\right)
-\frac{39\beta_{1,1}^{(2)}}{16}\right)+
x^4\left(\frac{29}{24}+4\alpha^{(2)}_{1,2}-
32\alpha^{(2)}_{1,3}\right)\right]
\end{eqnarray}
Notice that all terms proportional to $\omega^0$ and $\omega^1$ have
canceled out, which again shows that the lowest-order 
approximation for the
eigenvalue and eigenstate are already correct and agree with the
results for the continuous case. The three coefficients related to the
$\psi^{(2,2)}$ can be solved from the coefficients of $\omega^2$,
$\omega^3x^2$ and $\omega^4 x^4$ and they read
\begin{equation}
\alpha^{(2)}_{1,2}=-7/32,\quad \alpha^{(2)}_{2,2}=1/96,
\quad\mathrm{and}\quad  \beta^{(2)}_{1,1}=1/4.
\end{equation}
Substituting these into the set of equations and extending the calculation
to order $\omega^6$ we find the subsequent coefficients to be
\begin{equation}
\alpha^{(2)}_{1,3}=-275/1536,\quad\alpha^{(2)}_{2,3}=23/1024,
\quad \alpha^{(2)}_{3,3}=-1/1280,\quad\mathrm{and}\quad
\beta^{(2)}_{1,2}=37/256.
\end{equation}
After solving a sufficient number of coefficients
$\alpha^{(n)}_{kl}$ and $\beta^{(n)}_{kl}$ one should start
searching for regularities in the solution. 

Almost immediately we guessed the polynomial character of
$\alpha^{(n)}_{kl}$, first in terms of $n$ and later 
noticing that they should be written in terms of $\hat n$
as in Eq.~(\ref{eq:genalpha}). This considerably helps
solving the coefficients  $\beta^{(n)}_{kl}$ 
as for larger values of $n$ the
coefficients $\alpha^{(n)}_{kl}$ appear as constants,
not unknowns. 

In the beginning, we tried to solve all possible terms
up to a given order in $\omega$. First one should notice
that only terms with $l\le m$ are required for the solution
function $\psi_x^{(n,m)}$. Assuming that the previous orders 
have been explicitly obtained, means that only the equations
corresponding to $m'=m$ in Eq.~(\ref{eq:firstord})
have to be solved. In addition, generally known 
coefficients $\alpha_{k,m}^{(n)}$ identically satisfy 
equations corresponding to the highest powers of $\xi$.
Explicitly, if we assume that coefficients 
$\{\alpha_{k,m}^{(n)}\}_{k=k_0}^m$ are known, 
only the equations for  
\begin{equation}
\{\omega^m \xi^{n+2k_0-2l'}\}_{l'=0}^{k_0+k'}
\end{equation}
are required and the expansion of Eq.~(\ref{eq:expanded}) 
has to be carried out  up to and including the order
$\omega^{m+k_0+n/2}$ for coefficients $l\le m$.

After obtaining a rather complicated 
expression for the coefficients $\beta_{k,l=3}^{(n)}$, we 
happened to transform it into form equivalent to
the present form and  conjecture the general form of 
$\beta_{kl}^{(n)}$ in  Eq.~(\ref{eq:genbeta}). 
The most important lesson taught by the discretised
harmonic oscillator when solving the coefficients
is that your numbers may be wrong, but the general
forms usually are not. On several occasions, this 
became painfully obvious when the numbers did not 
check. Each and every time the general forms were
correct, but the used expansion of 
Eq.~(\ref{eq:expanded}) or the numbers inserted into it 
were not.

Later on, we started to study the regularities in the 
general expressions. The polynomial structure of the
coefficients $\alpha^{[l']}_{kl}$ that do not contain
any Gamma functions was relatively easy obtain, but
the other set required a real stroke of luck. We managed
to write some of these coefficients $\alpha^{[l']}_{kl}$ 
as explicit products. After being pointed out, by 
\textsc{Mathematica}, that the first two 
could be written in terms of Gamma functions, it was only
a question of finding the correct Gammas before 
Eq.~(\ref{eq:realgenalp}) was written. In order to
appreciate the technical part of obtaining the general
form of the coefficients we point out that the
coefficient $\alpha_{k,k+4}^{(n)}$
was completed by solving the $12^{\mathrm{th}}$ order solution 
$\psi_{n=0}^{(m=12,x_0)}$ and confirmed by the case $n=1$.
Further terms have been obtained by solving the asymptotical
differential equations~(\ref{eq:difftrun}).

The regularities in the coefficients $\{\beta_{kl}^{(n)}\}$
have been found out using by studying the expansions with
respect to $k$ and $k'$. By conjecturing the recurring
appearance of $(10k'-k)$ in Eqs.~(\ref{eq:kcommak}) and
(\ref{eq:kcommak2}) it becomes possible to solve the
quantities defined in 
Eqs.~(\ref{eq:betalead1})--(\ref{eq:betadiff2}). In addition
to these, the general expression for $\rho_{k,l-3}^{(l)}$
can be obtained from the known coefficients.

Finally, we will estimate the difficulty of obtaining
the explicit asymptotical solution $\psi_n^{(m,x_0)}$. We
assume that both the expansion of the eigenvalues up to the
required order and the solution $\psi_n^{(m-1,x_0)}$ have been obtained
in advance. The coefficients $\alpha^{(n)}_{km}$ can be determined 
from the exponential parts of the eigenvectors up to and 
including the case $n=m-1$. The completely general expressions
in Eq.~(\ref{eq:realgenalp}) are finished at much slower a pace.
The asymptotically satisfied differential equations~(\ref{eq:difftrun})
speed up this process considerably.

Obtaining the coefficients $\beta^{(n)}_{km}$ is more difficult.
The general form~(\ref{eq:genbeta}) shows that all states up
to $n=4m-3$ must be solved. The explicit expressions for the
leading parts, i.e.  $B(l,2l-2)$, $B(l,2l-3)$,  $\bar B(l,2l-3)$,
and  $\bar B(l,2l-4)$  make this task easier by 5 states. 
Thus all states up to $n=4l-8$ must be found, unless further
general properties are found.  

Regardless of these simplifications, the number of required terms 
and participating equations grows quite fast. Obviously, the
general form of the coefficients in the exponential factor is 
much easier to obtain and thus they should be applied as early 
as possible. It is also possible that considerable simplifications
or generalisations for the known coefficients lurk just around
the corner. This has already happened on several occasions
so far. We still choose to pause here, as the given 
general expressions have been validated rather 
convincingly and it not obvious,
how, if at all, the next orders in the expansion 
would improve the results qualitatively. We hope a
solid foundation has been laid for those striving 
towards the complete, asymptotical solution for 
the discretised harmonic oscillator.

\section{Proving the solution and some general properties
\label{sec:proof}}

Finally, we attack the difficult problem of actually showing that
the solution is a general one. Thus far we have solved the
equations for an increasing number of eigenstates using
Eq.~(\ref{eq:expanded}). This formulation is the best if
actual numerical values of the coefficients $\alpha_{kl}^{(n)}$ 
and $\beta_{kl}^{(n)}$ are sought after. This is explained
by symbolic math being most effective when the number of
unknowns and symbols is as small as possible. In principle,
the process explained below could be used for obtaining
recursion relations between the coefficients of the
solution and, subsequently, the full solution. 
Presently, we only show that the equations corresponding
to leading orders up to $\omega^3$ are satisfied identically.   

We  have to solve the equations corresponding to 
$\{\omega^m \xi^{n+m-2l'}\}_{l'=0}^{k'+m}$ in order to obtain
the $m^{\mathrm{th}}$ order solution. We have now obtained the
explicit solution up to the seventh order so we can check if it is 
correct. For this purpose, we must write Eq.~(\ref{eq:expanded})
explicitly in terms of $u:=\sqrt\omega$ and $\xi$, although
odd powers of $u$ eventually cancel. Multipliers of 
$\alpha_{kl}^{(n)}$ and $\beta_{kl}^{(n)}$  now read
\begin{equation}
u^{2(l-1)}[(\xi\pm u)^{2k}-\xi^{2k}]\quad\mathrm{and}\quad
(\xi\pm u)^{n+2(k-k')}, 
\end{equation}
respectively. On the right-hand-side the non-trivial term is given
by $\omega\xi^2/2$. Expanding all terms multiplying a fixed
term $h_k^{(n)}$ up to the order yields terms
\begin{equation}
h_k^{(n)}\left[1+u^2\left(2(k-k')-\hat n/2+\frac{\xi^2}2+\frac{(n-1 + 2(k -k'))
   (n + 2(k -k'))}{\xi^2}\right)\right]=h_k^{(n)}\left(1 + u^2\left(-\hat n/2 
    + \frac{\xi^2}2\right)\right).
\end{equation}
The terms proportional to $\xi^2$ cancel and equating each power of 
$\xi$ separately yields an equation 
\begin{equation}
h_k^{(n)}2(k'-k)+[2(k+1)^2\mp(k+1)]h_{k+1}^{(n)}=0,
\end{equation}
where the signs $+$ and $-$ corresponds the even and odd values
of $n$, respectively. The above equation is identically satisfied
by the Hermite polynomials, which proves that the 
first order solution $\psi_n^{(1,x_0)}$ is correct.
A careful observer immediately asks about the second order 
corrections $\beta^{(n)}_{k,2}$ which also yield terms proportional
to $u^2$. However, these coefficients are not fixed at all by 
Eq.~(\ref{eq:expanded}) in the order $u^2$. The only term that is
easily solvable from this relation in the dominant coefficient
$\alpha_{11}^{(n)}=-1/2$ which removes $h_{k-1}^{(n)}$ from the
recursion relations. Later on, the dominant coefficients 
$\{\alpha_{kk}^{(n)}\}_{k=1}^m$ cancel the term 
$h_{k-m}^{(n)}$ in the order $\omega^m$.

In the next order $\omega^2$ we insert the solved coefficients and obtain
for even values of a recursion relation 
\begin{eqnarray}
&&6(n+2 -2 k)h_{k-1}^{(n)}+[(2k^3 + k^2(42 - 11n) - 6n - 3n^2 - 
  k(-6 + 9n - 5n^2)]h_{k}^{(n)}-\cr
&&\, (1 + k)(1 + 2k)(22 + 31k + k^2 - 5n - 5kn)h_{k+1}^{(n)}+
    2(2k + 4)(2k + 3)(2k + 2)(2k + 1)h_{k+2}^{(n)}=0,
\end{eqnarray}
which is again identically satisfied by the Hermite polynomials.
For odd values of $n$, we find a similar recursion relation, once
we replace $k'=n/2$ by $k'=(n-1)/2$. This 
completes the proof in order $\omega^2$ and validates 
the second-order eigenvectors $\psi_n^{(2,x_0)}$.

In the third order the recursion relation for even values of $n$ reads
\begin{eqnarray}
&& 180 (-4 + 2 k - n)h_{k-2}^{(n)}+30(62 - 42k - 24k^2 + 4k^3 + 74n - 10k n -
    22k^2n +  17n^2 + 10kn^2)h_{k-1}^{(n)}+\cr
&&  [- 450n - 450n^2 - 90n^3 -10k^5 + 
    k^4(-452+ 105 n) + 
    k^3(-2332 + 2458n - 300n^2) + \cr
&&  k^2(4230 + 4912n - 1204n^2 + 125n^3) + 
    k(-300 - 585n - 1258n^2 + 179n^3)]h_{k}^{(n)}+\cr
&&(1 + k)(1 + 2 k)(1022 + 2705k + 3684k^2 + 326k^3 + 
        5k^4 - 2274 n - 4430k n - 1726k^2 n -50k^3 n + 454 n^2 +\cr
&& \,  579k n^2 + 125k^2 n^2)h_{k+1}^{(n)}-
16(1 + k)(2 + k)(1 + 2k)(3 + 2k)
		  (110 + 101k + 5k^2 - 50n - 25k n)h_{k+2}^{(n)}+\cr
&&256 (1 + k) (2 + k) (3 + k) (1 + 2 k) (3 + 2 k) (5 + 2 k)h_{k+3}^{(n)}=0.
\end{eqnarray}
Because the Hermite polynomials satisfy this and the corresponding
relation for odd values of $n$ the solution $\psi_n^{(3,x_0)}$
has been rigorously proven as correct.

The eigenvectors $\psi_n^{(m,x_0)}$ tend to the 
eigenvectors $\psi_n^{(x_0)}$ of $H(x_0)$ at an asymptotical
rate proportional to $\omega^m$. The exact eigenvectors
are orthogonal as eigenvectors of a Hermitian matrix and 
by their closure relation we can write
\begin{equation}
\psi_n^{(m,x_0)}\sim \psi_n^{(x_0)}+\omega^m\sum_{n'}b_{n'}
\psi_{n'}^{(x_0)},
\end{equation}
where $b_{n'}$ are finite constants such that $\sum_{n}\vert b_n\vert^2<
\infty$ in the limit $\omega\rightarrow 0$.
The orthonormality relation for the asymptotical solutions thus reads
\begin{equation}
\langle \psi_n^{(m,x_0)}\vert\psi_{n'}^{(m,x_0)}\rangle=\delta_{nn'}+
\mathcal{O}(\omega^m),
\end{equation}
provided that the sum $\sum_n \vert b_n\vert$ is finite for both states.
In other words, the eigenvectors $\psi_n^{(m,x_0)}$ become orthonormal
at the rate of $\omega^m$. Numerical checks seem to confirm this, at 
least for relatively small values of $n$. 

As a final effort, we outline a plausible "proof" for the asymptotical
convergence. As a first step, we show that without loss
of generality we can examine a finite truncation of the 
eigenvector $\psi_n^{(n,x_0)}$, the vector 
$\psi_{n,j_0}^{(x_0)}:=
\{\psi_{j-x_0}^{(n)}\}_{j=-j_0}^{j_0}$ for sufficiently large $j_0$. 
For sufficiently large values of $\vert j\vert$ the eigenvalue 
$\lambda_n$ becomes insignificant in Eq.~(\ref{eq:general}) and 
we write an approximate equation   
\begin{equation}
(\psi_{x-1}^{(n)}-2\psi_x^{(n)}+\psi_{x+1}^{(n)})/(x^2)=
\omega^2.
\end{equation}
For sufficiently large values of $x$ and/or $j$ the sign of $\psi_x^{(n)}$
is constant and this equation shows that the function 
$\tilde \psi_x=\psi_{j_0-x_0}^{(n)}\exp(-\omega(x^2-j_0^2)/(2+\varepsilon))$,
for some small $\varepsilon>0$, is a dominant sequence for $\psi_x^{(n)}$. 
Now, the limiting sequence of  norms
\begin{equation}
\lim_{j_0\rightarrow \infty}\Vert \psi_{n}^{(x_0)}-\psi_{n,j_0}^{(x_0)}\Vert 
\end{equation}
vanishes exponentially with respect to $j_0$. In other words, we can 
always find a finite $j_0$ such that the error in the norm is 
sufficiently small. 

Next we use the fact that solution $\psi_{n}^{(m,x_0)}$ satisfies 
the eigenvalue equation~(\ref{eq:general}) up to the order $\omega^m$
when written in terms of $\xi$. Thus we can write 
\begin{equation}
\frac{\psi_{x-1}^{(n,m)}+\psi_{x+1}^{(n,m)}}{2
\psi_x^{(n,m)}(-\lambda_n+\omega \xi^2/2)}=1+\mathcal{O}(\omega^{m+1}).
\label{eq:error}
\end{equation}
We fix the scales of the eigenvectors by setting
$(\psi_{n}^{(m,x_0)})_{j}=(\psi_{n}^{(x_0)})_{j}$ for an arbitrary $j$. 
It would be very tempting to say that Eq.~(\ref{eq:error}) 
implies $(\psi_{n}^{(m,x_0)})_{j+1}=(\psi_{n}^{(x_0)})_{j+1}
(1+\mathcal{O}(\omega^{m+1}))$ and then wonder why convergence 
is not asymptotically proportional to $\omega^{m+1}$. 
As already explained the solution $\psi_{n}^{(m,x_0)}$ does not
fix the coefficients $\beta_{k,m+1}^{(n)}$ which most definitely
yield terms proportional to $\omega^m$. Thus we obtain a relation
\begin{equation}
(\psi_{n}^{(m,x_0)})_{j+1}=(\psi_{n}^{(x_0)})_{j+1}
(1+\mathcal{O}(\omega^{m})).\label{eq:consequtive}
\end{equation}
By matching the eigenvectors at $j=0$, and expanding the 
components to the finite values $\pm j_0$ shows that the
order of error is $\omega^m$ for all components with $\vert j\vert \le j_0$. 
Becase the  error caused by the truncation is insignificant the result
holds for the full eigenvectors and we obtain the desired result 
\begin{equation}
\Vert \psi_n^{(m,x_0)}-\psi_n^{(x_0)}\Vert\sim \omega^m,
\end{equation}
or at least show that the result is quite plausible.

\section{Conclusions\label{sec:conclu}}

We have obtained an explicit, asymptotical solution for
the discretised harmonic oscillator. Both the eigenvalues and
eigenvectors have been obtained and we can choose a prespecified
rate of convergence towards the exact solutions. This is done by
truncating the ansatz solution accordingly. Because the
problem can be mapped onto the Mathieu differential equation,
we simultaneously provide asymptotical expressions 
for the Mathieu functions. The Schr\"odinger equation of the 
quantum pendulum corresponds to the Mathieu equation,
which yields immediate applications for the results.

The method described above can be generalised to accommodate 
several coordinate dimensions with only minor changes. This should
make the results of Ref.~\onlinecite{aun02} both
more transparent and more rigorous. The tunnelling-charging
Hamiltonian of a Cooper pair pump corresponds to a
modified multi-dimensional Mathieu equation.

Alternatively, ansatzes similar to Eq.~(\ref{eq:genansatz})
could be constructed in case of difference equations that
become identical to analytically solvable differential
equations in some asymptotical limit. Initially, the
problem assumes the form of an infinite-dimensional,  
two-parameter [eigenvalue] problem, where the asymptotical solutions
[eigenvalues and  eigenvectors]  must be obtained. 
The ansatz maps the problem onto an infinite set of 
algebraic equations that must solved. If the form of the ansatz is correct, 
one may determine some general properties of the exact solution.

\acknowledgments{
This work has been supported by the Academy of Finland
under the Finnish Centre of Excellence Programme 2000-2005
(Project No. 44875, Nuclear and Condensed Matter Programme at JYFL).
Dr.~L.~Kahanp\"a\"a is acknowledged for insightful discussions
and suggestions. The author thanks Prof.~J.~Timonen for valuable
references added to the final draft of the manuscript.}

\bibliography{Hermite}

\end{document}